\documentstyle[12pt]{article}
\def\appendix#1{
  \addtocounter{section}{1}
  \setcounter{equation}{0}
  \renewcommand{\thesection}{\Alph{section}}
  \section*{Appendix \thesection\protect\indent #1}
  \addcontentsline{toc}{section}{Appendix \thesection\ \ \ #1}
  }
\def\tr#1{\,{\rm tr}\,#1\,}

\def\iint{\int\hspace{-.6em}\int}

\def\eop{\vspace*{\fill}\pagebreak}
\def\be{\begin{equation}}
\def\ee{\end{equation}}
\def\bea{\begin{eqnarray}}
\def\eea{\end{eqnarray}}

\def\Re#1{\,{\rm Re}\,#1}
\def\Im#1{\,{\rm Im}\,#1}
\def\curraddr#1\endcurraddr{\address {\it Current address\,}: #1\endaddress}
\def\qqq#1{\sqrt{\frac {#1^2}{4}+b#1+c}}
\def\pd#1#2{\frac{\partial #1}{\partial #2}} %partial derivative
\def\pdd#1#2{\frac{\partial ^2{}}{\partial #1 \partial #2}} %double derivative
\def\aaa{\tilde\alpha}
\def\L{\Lambda}
\def\h{\eta}

\title{{\bf \mbox{} \\THE MULTICRITICAL KONTSEVICH--PENNER MODEL}\vspace{.5cm}}
\author{{\bf L. Chekhov}\thanks{E--mail: \ chekhov@qft.mian.su}
\date{ }
\vspace{.5cm} \\
{\it Steklov Mathematical Institute} \\
{\it Vavilov st.42, GSP-1, 117966 Moscow, RUSSIA}\\ \\
and \\ \\
{\bf Yu. Makeenko}\thanks{E--mail: \ MAKEENKO@nbivax.nbi.dk \ \ / \ \
MAKEENKO@desyvax.bitnet}
\vspace{.5cm} \\{\it The Niels Bohr Institute} \\
{\it Blegdamsvej 17, DK-2100 Copenhagen, DENMARK} \\
{\it and} \\
{\it Institute of Theoretical and Experimental Physics} \\
{\it B.Cheremuskinskaya 25, 117259 Moscow, RUSSIA}}

\begin{document}

\maketitle

\vspace{-15cm}

\begin{flushright}
NBI-HE-92-03 \\ January 13, 1992
\end{flushright}

\vspace{14cm}

\begin{abstract}
We consider the hermitian matrix model with an external field entering the
quadratic term $\tr{(\Lambda X\Lambda X)}$ and Penner--like interaction term
$\alpha N(\log{(1+X)} -X)$.
An explicit solution in the leading order in $N$ is
presented. The critical behaviour is given by the second derivative of the
free energy in $\alpha$ which appears to be a pure logarithm, that
is a feature of $c=1$ theories. Various critical regimes are possible,
some of them corresponds to critical points of the usual Penner model, but
there exists an infinite set of multi-critical points which differ by
values of scaling dimensions of proper conformal operators.
Their correlators with the puncture operator are given in genus zero by
Legendre polynomials whose argument is determined by an analog of the
string equation.
\end{abstract}

\vspace{.5cm}
\noindent
Submitted to {\sl Modern Physics Letters A}

\eop

\section {Introduction}

Recently the external field problem for matrix models has achieved a
remarkable interest due to the work by Kontsevich \cite{Kon91} who
had represented the partition function of 2D topological gravity as that of a
hermitian one-matrix model in an external field. Along this line, a direct
proof of the Witten's conjecture \cite{Wit90} about an equivalence of 2D
topological and quantum gravities has been obtained \cite{Wit91,MMM91,GN91b}.

On the other hand, the Kontsevich matrix model can be explicitly solved
\cite{KK89,GN91,MS91} order by order of genus
expansion making use of the standard methods which were
first introduced for an analogous unitary matrix problem
\cite{BG80}. This solution has been applied recently by Itzykson
and Zuber \cite{IZ92} to calculate explicitly the intersection indices on the
moduli space.

The original Kontsevich model is reduced by a linear shift of an integration
variable to a generic form
\be
{\cal F}[\Lambda] = \int DX\, \hbox{e}^{\,N\tr{\left(\Lambda
X+V(X)\right)}}
\label{1.1}
\ee
with a cubic potential $V(X)$ and the integral going over $N\times N$
hermitian
matrices $X$. For an arbitrary $V(X)$, the partition function
(\ref{1.1}) is
associated with the so-called generalized Kontsevich model. As is advocated by
Kharchev et al.\ \cite{KMMMZ91},
this model interpolates between arbitrary
$(p,q)$ double-scaling limits of the standard multi-matrix model. It is
crucial for applications that the exponent in (\ref{1.1}) depends on the
external field $\Lambda$ linearly.

One might think, however, about another way of incorporating the external
field by modifying the quadratic term:
\be
{\cal Z}[\Lambda] = \int DX\, \hbox{e}^{\,N\tr{
\left(-{1\over 2}\Lambda X \Lambda X+V(X)\right)}} .
\label{1.2}
\ee
A model of this kind was first introduced by Das et al.\ \cite{DDSW90}.
In the language of discretized random surfaces, it corresponds to an external
field which is connected to curvature of the world sheet.

One may ask a question: which form of the potential $V(X)$ in (\ref{1.2})
permits  such changing the variables that results  in an integral of the type
(\ref{1.1})? The only nontrivial choice is the nonpolynomial
$V(X)=\alpha (\log{(1+X)}-X)$ which is associated with the Penner model
\cite{Pen86}. As is pointed out by Distler and Vafa \cite{DV91}, the
double-scaling limit of the Penner model corresponds to $c=1$ CFT or to the
$d=1$ string at the self dual radius $R=1$ which is first solved by Gross and
Klebanov \cite{GK90}. Further studies of generalizations of the Penner
model at multi-critical points have been done recently in
\cite{Tan91,CDL91}.

We solve the model (\ref{1.2}) in genus zero explicitly using the method of
Br\'ezin and Gross \cite{BG80}. We show that the second derivative of
$\log{\cal Z} $ in $\alpha$ is a pure logarithm which is a feature of $c=1$
theories. We study various multi-critical regimes and relate the traces of
inverse powers of $\Lambda$ (the Kontsevich--Miwa variables) to sources for
conformal operators. We calculate the correlation functions of the
conformal operators with the puncture operator and show that it is given
by Legendre polynomials whose argument is determined by an analog of the
string equation which is in our case a set of two equations for two
variables. We calculate the proper anomalous dimensions which
turn out to be different at different multi-critical points.

\section{Formulation of the model}

\setcounter{equation}{0}
We consider the hermitian one-matrix model in an external field:
\be
{\cal Z}[\L]=
\int DX\,\hbox{e}^{\,N\,\tr{\left\{-\frac 12\L X\L X+\alpha\bigl[\log
(1+X)-X\bigr]\right\}}},\qquad \L = \hbox{diag\,}(\L_1,\ldots\,,\L_N).
\label{2.1}
\ee
Our strategy to deal with the model (\ref{2.1}) is to reduce it to the
Kontsevich type integral (\ref{1.1}). This can be done by a substitution
$\L ^{1/2}X\L ^{1/2}\rightarrow X$. We obtain
\bea
{\cal Z}[\L]&=&
\hbox{e}^{-\frac N2 \,\tr{\L ^2}+\alpha N^2}(\det \L )^{-N(\alpha
+1)}\times \nonumber \\
 & &\int DX\,\hbox{e}^{\,N\,\tr{\left(-{X^2\over 2}+\h X\right)}}
(\det X)^{\alpha N},
\qquad \h = \L -\alpha \L ^{-1}.
\label{2.3}
\eea

Now our goal is to study the integral
\be
Z[\h]=
\int DX\,\hbox{e}^{-N\,\tr{\left\{\frac 12 X^2-\h X\right\}}} (\det X)^{\alpha
N}.
\label{2.4}
\ee
There are different techniques to proceed with this integral. In the next
sections we shall treat it using the method \cite{BG80} of
Schwinger--Dyson equations written in terms of eigenvalus, which has
been applied to hermitian one-matrix models in \cite{KK89,GN91,MS91,MMM91},
and discuss Virasoro constraints for this model.

We conclude this section with the Itzykson--Zuber--Mehta technique
\cite{IZ80}
for integration over angular variables in multi-matrix models. We assume that
it is well known and
do not intend to engage into any detailed description. In terms of eigenvalues
of the matrices $X$ and $\h$, (\ref{2.4}) can be expressed as follows:
\be
Z[\h]={\cal Z}\int_{-\infty}^{\infty}\,\prod_{i=1}^{N}x_{i}^{\alpha N}\,dx_i
{\Delta [x] \over\Delta [\h ]}\,\exp \left\{-N \sum_{i=1}^{N}\left(
{x_{i}^{2} \over 2}-\h _{i}x_i\right)\right\},
\label{2.5}
\ee
where $\Delta [x] = \prod_{i>j}^{N}(x_i-x_j)$ is the Van der Monde
determinant. In what follows we assume that $\alpha N$ is positive integer.
Using the relation
\be
\int_{-\infty}^{\infty}\,dx\,x^n\,\hbox{e}^{-x^2/2+\h x}=i^{-n}\,H_n(i\h )
e^{\h ^2/2}
\ee
we may integrate out all $x_i$'s in the Eq.(\ref{2.5}):
\be
Z[\h]={\cal Z}N^{\alpha N^2/2+N(N+1)/2}\Delta ^{-1}[\h ]
\det_{ {1\le k\le N}\atop { 1\le j\le N}} \left\Vert i^{-\alpha N-(k-1)}
H_{\alpha N+k-1}(i \sqrt N \h _j)\right\Vert \exp \biggl(N\sum_{i=1}^{N}
{\h _i^2 \over 2}\biggr).
\label{2.6}
\ee
Here $H_n(i\h )$ are the Hermite polynomials of an imaginary argument.
As is shown by Kharchev et al.\ \cite{KMMMZ91}, the determinant formula
(\ref{2.6}) implies $Z[\h]$ to be a $\tau$-function.

It is clear from Eq.(\ref{2.6}) that $Z[\h ]$ is a symmetric polynomial of
degree
$\alpha N^2$ in variables $\h _i$. But it is still unclear how to deal
further with the determinant. More effective way to obtain the answer is to
use the Schwinger--Dyson equations generating Virasoro constraints and the
``master equation'' in this model.

\section{Virasoro constraints}

\setcounter{equation}{0}
The Schwinger--Dyson equation for our model (\ref{2.1}) follows from
an invariance of $DX$ under ($X$-independent) variations of $X$:
\be
\int DX\,{\delta \over \delta X_{ij}}\,\hbox{e}^{\,-N\,\tr{\left(\frac
{X^2}{2}-\h X -\alpha \log X\right)}}=
\left\langle {\delta \over \delta X_{ij}}\,\tr{\left(\frac {X^2}{2}-
\h X -\alpha \log X\right)}\right\rangle =0.
\label{3.1}
\ee

Taking into account that $N<x_{ij}F>={\delta \over \delta \h _{ij}}\,<F>$ and
applying one additional external derivative over $\h $ to eliminate the
nonlocal contribution arising from the variation of $\alpha \log X$, we have
\be
\left[ -\pdd {\h _{ij}}{\h _{jk}}+N\left(\h \pd {}{\h }\right)_{ik} +
(1+\alpha )N^2\delta _{ik}\right]\,Z[\h ]=0.
\label{3.2}
\ee
This equation is discussed in detail below.

One could make alternatively the $X$-dependent variation $\delta X =
\epsilon_n X^{n+1} $ which would immediately lead for $n\geq 0$
to a set of Virasoro constraints of the type advocated by
Semenoff and one of the authors \cite{MS91}
for the model (\ref{1.1}) with arbitrary polynomial $V(X)$:
\begin{equation}
{\cal L}_n Z[\h] =0     \hbox{  \ \ for \ \ \ } n\geq 0
\label{calconstr}
\end{equation}
and
\begin{equation}
{\cal L}_n= \sum _i \left( - (\nabla _i)
^{n+2} + N(\nabla _i) ^{n+1} \h _i + \alpha N (\nabla_i)^n+
{1 \over 2} \sum _{k=0} ^{n} \sum _{j \neq i} (\nabla _i)^k (\nabla
_j)^{n-k}
\right)
\label{calL_n}
\end{equation}
where
\begin{equation}
\nabla _i
= {\partial\over \partial \h_i} + \sum_{j \neq i}{1 \over \h_i-\h_j}
\label{nabla}.
\end{equation}
However, the $n=-1$ equation which corresponds in the polynomial case to
${\cal L}_{-1}$ operator is now nonlocal. This seems to be a
reflection of the fact that we are dealing with a $c\neq 0$ theory.

Let us now turn to Eq.(\ref{3.2})
which resembles in many details the equation \linebreak
$\left(\delta^2/\delta \h ^2+\h \right)Z[\h ]=0$
generating \cite{MMM91,GN91b} the continuum
Virasoro constraints in the standard Kontsevich model.
For our case we do not know in advance proper expressions for continuum
time-variables in terms of the eigenvalues of $\L$. We show below by
explicitly solving the model (\ref{2.1}) in genus zero that those are given by
the Kontsevich--Miwa transformation
while the problem of constructing continuum Virasoro (or W-type)
constraints for our model is beyond the scope of present publication.

However, it is worth mentioning that for the model (\ref{2.1}) there
exists some Virasoro algebra in terms of ``reverse times''
$q_k = {1\over kN}\,\tr{\h ^k}$ while the standard Kontsevich--Miwa times
are defined via expansion in negative powers of the matrix $\h $.

Doing the contraction of Eq.(\ref{3.2}) with $(\h ^n)_{ki}$ we obtain the set
of
constraints:
\be
\sum_{s=2}^{\infty} q_{s+n-2}L_s\,Z[\h ],
\label{3.3}
\ee
where
\be
L_s= \sum_{a+b=s}\pdd {q_a}{q_b}+\sum_{b=1}^{\infty}bq_b\pd {}{q_{s+b}}+
\pd {}{q_s} +\pd {}{q_{s-2}}-(1+\alpha )\delta _{s,2}.
\label{3.4}
\ee
One may prove that for finite $N$ these conditions are sufficient to fix
uniquely the form of the solution which is the polynomial of the maximum
order $\alpha N^2$ in $q_i$ multiplied by a trivial Gaussian factor.
The continuous limit is rather nontrivial
and will be discussed elsewhere.

We shall consider in what follows not the Eq.(\ref{3.2}) but the equation for
the ratio of the determinants (\ref{2.6}). $Z[\h ]$ differs from it by a
factor
$\hbox{e}^{\frac N2 \,\tr{\h ^2}}$.
Pulling this factor through Eq.(\ref{3.2}), we have
\be
\left[ -\left(\pdd {\h }{\h }\right)_{ik}-N\left(\h \pd {}{\h }\right)_{ik}+
\alpha N^2\delta _{ik}\right]\,
{\det _{(k,j)} \left\Vert
H_{\alpha N+k-1}(i \sqrt N \h _j)\right\Vert \over \Delta [\h ]}=0.
\label{3.5}
\ee

There also exists an interesting
representation of these constraints not in Kontsevich--Miwa but rather
Macdonald's like variables \cite{Mac}:
\be
g_k = \sum_{\{i_1,\,\ldots,i_k\}} \h _{i_1}\cdots\h _{i_k},
\label{3.6}
\ee
where the sum runs over all products of $k$ different $\h _i$'s. Then the
symmetric function in Eq.(\ref{3.5}) is the polynomial of degree $\alpha
N$
(not $\alpha N^2$) in the variables $g_k$, the higher term being
$g_N^{\alpha N}$. Moreover, the coefficient standing by term
$g_{N-k_1}\cdots g_{N-k_{\alpha N}}$ depends only on the set $\{k_1,\ldots,
k_{\alpha N}\}$ but not only $N$ itself. In this case the explicit solutions
for few lowest values of $\alpha N$ can be found.

\section{The ``master equation''}

\setcounter{equation}{0}
In this section we rewrite Eq.(\ref{3.2}) (or (\ref{3.5})) in terms of
eigenvalues of the matrix $\h$. We obtain an equation (the ``master
equation'') which
is similar to that of Refs.\cite{GN91,MS91,MMM91} for the Kontsevich model.

We want to find a suitable form
of Eq.(\ref{3.5}) in terms of eigenvalues of the matrix $\h$:
\be
\h =UHU^+,\qquad H=\,\hbox{diag\,}\,(\h _1,\ldots, \h _N),
\label{4.1}
\ee
where $U$ is a unitary matrix. We contract Eq.(\ref{3.5}) with the special
matrix
$F^{(j)}$ of the form:
\be
F^{(j)}=Uf_jU^+, \qquad f_j=\,\hbox{diag\,}\,(0,\ldots,0,1_{(j)},0,\ldots,0).
\label{4.2}
\ee
Acting on arbitrary monom $t_{a_1}\cdots t_{a_s}$ composed from the times
$t_{a_i}=\tr{\h ^{a_i}}$, where $a_i$ are no more restricted to be nonnegative
integers, we have
\be
\tr{\left( F^{(j)}\h \pd {}{\h }\right)}=\h _j\,\pd {}{\h _j},
\label{4.3}
\ee
and for the first term:
\bea
\tr{\left\{(Uf_jU^{+})_{ks}\,\pdd {\h _{ts}}{\h _{kt}}\right\}}\,t_{a_1}\cdots
t_{a_s} = \nonumber \\
=(Uf_jU^{+})_{ks} \pd {}{\h _{ts}}\left\{ \sum_{l=1}^{s} a_l(\h ^{a_l-1})_{tk}
t_{a_1}\cdots \hat t_{a_l} \cdots t_{a_s}\right\}= \nonumber \\
=(Uf_jU^{+})_{ks} \left\{ \sum_{{l,p=1}\atop {l\ne p}}^{s} a_l a_p
(\h ^{a_l-1})_{tk}(\h ^{a_p-1})_{st}
t_{a_1}\cdots \hat t_{a_l} \cdots \hat t_{a_p} \cdots
t_{a_s}+\right. \nonumber \\
+ \left.\sum_{{l=1}\atop \mbox{}}^{s} a_l
\left( \pd {}{\h _{ts}}(\h ^{a_l-1})_{tk}\right)
t_{a_1}\cdots \hat t_{a_l} \cdots t_{a_s}\right\}=
\left\{ {\left( \pd {}{\h _j}\right)}^2 +\sum^{}_{i \ne j}{\partial _j-
\partial _i \over \h _j -\h _i}\right\}\,t_{a_1}\cdots t_{a_s}.
\label{4.4}
\eea

Thus we gave the direct proof of the ``master equation'' in our theory:
\be
\left\{ \partial _j^2  +\sum^{}_{i \ne j}{\partial _j-
\partial _i \over \h _j -\h _i}+N\h _j\partial _j -\alpha N^2\right\}\,
\tilde Z[\h ]=0,
\label{4.5}
\ee
where $\tilde Z$ stands for the ratio of the determinants (\ref{2.6}).

\section{The genus zero solution}

\setcounter{equation}{0}
Now we concentrate on the solution to Eq.(\ref{4.5}) in the spherical
(or genus zero) limit $N \rightarrow \infty$.
In order to describe the solution of Eq.(\ref{4.5}) at large $N$, let
us define the effective potential
\be
W[\h ] \equiv {1 \over N^2} \log{Z[\h ]}
\label{4.6}
\ee
which is normalized to be O(1) as $N\rightarrow\infty$.
The method of solving Eq.(\ref{4.5}) in the large-$N$ limit is described
in the Appendix. The result for the initial $W[\h ]$ reads
\bea
W[\h ] = \frac 12 \left( \aaa -\frac 12\right)^2\log (c-b^2)
-\frac 52 b^2c -\aaa c +\frac {c^2}{4} +3\aaa b^2 +\frac 94 b^4+ \nonumber \\
+ \frac 1N \sum _i  \left[\frac 14 \h_i^2+ \left(\frac {\h_i}{2}
-b\right)\qqq {\h_i}
+\aaa \log \left(\h_i+2b+\sqrt{\h_i^2+4b\h_i+4c}\,\, \right)\right]
- \nonumber \\
- {1 \over 4N^2} \sum _{i,j}
\log{\left[{\,\h_i\h_j \over 4} +\frac b2 (\h_i+\h_j) +c+\qqq {\h_i}
\,\, \qqq {\h_j} \,\,\,\right]}.
\label{4.8}
\eea
The variables $b$ and $c$ are functionals of $\rho $ and $\aaa $. They are
determined by the nonlinear constraints:
\be
b+\frac {1}{4N} \sum_i {1\over \qqq {\h_i}}=0,
\label{4.9}
\ee
and
\be
\frac {1}{4N} \sum_i {\h_i \over \qqq {\h_i}}+c-3b^2=\tilde\alpha .
\label{4.10}
\ee
Due to these conditions the expression Eq.(\ref{4.8}) is stationary w.r.t. the
differentiation over $b$ and $c$.

Among others, Eq.(\ref{4.8}) possesses a remarkable property. If we take the
double derivative of $W[\h ]$ over $\aaa $, then, again due to Eqs.(\ref{4.9})
and (\ref{4.10}), we get
\be
{d^2 \over d\alpha ^2}W[\h ]=\log (b^2-c),
\label{4.11}
\ee
and the result coincides with that of applying the partial derivative
$\partial^2/\partial\alpha^2$ to $W[\h]$.
It is worth to note
that the answer is a pure logarithm. We shall show in a moment
that the logarithm is immediately connected to the logarithmic scaling
violation for $c=1$ models.

To deal with Eqs.(\ref{4.9}), (\ref{4.10}), let us expand the l.h.s.'s in
$1/\h$. Using the relation
\begin{equation}
{1 \over \sqrt {1-2xy+y^2}}=\sum _{n=0}^{\infty}P_n(x)y^n,
\end{equation}
where $P_n(x)$ are the standard Legendre polynomials which are
normalized by $P_n(1)= 1$,
we rewrite Eq.(\ref{4.9}) as
\be
\sum _{n=1}^{\infty}t_n P_{n-1}\left(-\frac {b}{\sqrt c}\right) \left( 2
\sqrt c \right)^{n-1} = 0
\label{5.6}
\ee
and Eq.(\ref{4.10}) as
\be
\sum _{n=1}^{\infty}t_n P_{n}\left(-\frac {b}{\sqrt c}\right) \left( 2
\sqrt c \right)^{n} = 2\alpha
\label{5.7}
\ee
where
\be
t_n={1\over N} \sum_i {1\over \h _i^n} - \delta_{n2} \hbox{\ \ \
\ \ \ } n\geq 1\,\,.
\label{5.8}
\ee

Some comments are now in order.
The transformation from the variables $\h_i$ to $t_n$ given by
Eq.(\ref{5.8}) is that of the Kontsevich--Miwa type for the
(generalized) Kontsevich model. The variables $t_n$ become independent
as $N\rightarrow\infty$ which is just our case. The role of $t_0$,
the cosmological constant, is now
played by $-\alpha$. Eqs.(\ref{5.6}), (\ref{5.7}) resemble the
corresponding equation for the Kontsevich model \cite{MS91},
which is in that case
nothing but the genus zero string equation, while we have now
{\it two} variables $b$ and $c$. The extra shift of $t_2$ in Eq.(\ref{5.8}) is
similar to that in the Kontsevich model (or in 2D topological gravity)
where it is associated with a perturbative background. Eqs.(\ref{5.6}),
(\ref{5.7}) can easily be solved in the following cases:
\begin{description}
\item[(i)] For $\h_i\rightarrow\infty$ when $t_2=-1$, $t_1=t_3=\ldots=0$.
The solution is $b=0,c=\alpha$ so that
Eq.(\ref{4.11}) gives $\log{(-\alpha})$ which recovers the known
solution of the Penner model in genus zero.
\item[(ii)] If all $\h_i$'s are
constant: $\h_i \equiv s$. One obtains the one-cut solution of the
generalized Penner
model with the Gaussian term added \cite{Tan91,CDL91}. In this case
Eqs.(\ref{4.9}) and (\ref{4.10}) lead to the relations
\begin{equation}
c-b^2=\left({1\over 4b}-b-\,\frac{s}{2}\right)
\left({1\over 4b}+b+\,\frac{s}{2}\right)
\label{xxx}
\end{equation}
and
\begin{equation}
\alpha =\left({1\over 4b}-3b-\,\frac{s}{2}\right)
\left({1\over 4b}+b+\,\frac{s}{2}\right).
\label{yyy}
\end{equation}
There are two possible singular points: the first is $\alpha =0$ resulting
from $1/4b+b+s/2=0$, and it is the auxiliary singularity of the Penner
model. The second possibility is $1/4b-b-s/2=0$, and it immediately gives
us $\alpha=\,-1$, without any reference to the value of $s$. It is just
the true  nonperturbative critical point of the standard Penner model.
\item[(iii)] For a pure Gaussian case $\alpha=0$.
One gets $b=-\sqrt c$ (remember that $P_n(1)=1$).
\end{description}
One can construct
perturbations of these models by finding the corresponding
perturbative solutions of Eqs.(\ref{5.6}), (\ref{5.7}) around the
discussed ones. One should not take the double-scaling limit of the
model (\ref{2.1}). Similarly to the (generalized) Kontsevich model, it
describes already a {\it continuum}\/ case.

An example of such a perturbation of the Gaussian case (iii) by the
Penner action can be obtained for $\alpha\rightarrow 0$.
Let $b+\sqrt c \sim \alpha$. Eq.(\ref{5.6}) determines then $\sqrt c$ versus
$\{t_n\}$:
\be
\sum _{n=1}^{\infty}t_n \left( 2\sqrt c \right)^{n-1} = 0,
\label{5.9}
\ee
while Eq.(\ref{5.7}) gives $b+\sqrt c$ versus $\alpha$:
\be
\left(b+\sqrt c\right)\sum _{n=1}^{\infty} n(n+1) t_n
\left( 2\sqrt c \right)^{n-1} = -2\alpha
\label{5.10}
\ee
(remember that $P_n^\prime(1)=n(n+2)/2)$. Eq.(\ref{4.11}) shows again
a logarithmic dependence on~$\alpha$.

Let us discuss now an interesting  nonperturbative solutions to
Eqs.(\ref{5.6}),
(\ref{5.7}). It looks similar to the multi-critical
one-cut solution of the hermitian one-matrix model with a polynomial
potential, where for the $K^{th}$ multi-critical point one puts all $t_n=0$
except $n=0,K$. Let us take $t_K\neq 0, t_1=\ldots=t_{K-1}=t_{K+1}
=\ldots=0$.
In this case the solution to Eq.(\ref{5.6}) reads
\begin{equation}
-{b \over \sqrt {c}}= x_j^{(K-1)},
\label{4.15}
\end{equation}
where $-1<x_j^{(K-1)}<1$ is $j^{th}$ root of equation $P_{K-1}(x)=0$ (every
$P_n(x)$ has exactly $n$ zeros on the interval $[-1,1]$). Let us denote
$p_j^{(K)}= P_{K}(x_j^{(K-1)})$, all $p_j^{(K)}$ being nonzero. Then for
$c$ and $b^2-c$ we have
\begin{equation}
c={1\over 4}{\left[ {2\alpha \over t_K p_j^{(K)}}\right]}^{2/K},
\label{4.16}
\end{equation}
\begin{equation}
b^2-c={1\over 4}\bigl[ {(x _j^{(K-1)})}^2-1 \bigr]{\left[ {2\alpha \over
t_K p_j^{(K)}}\right]}^{2/K}.
\label{4.17}
\end{equation}
One substitutes this solution into Eq.(\ref{4.11}) to see
a logarithmic dependence of $W$ on $\alpha$. Therefore we conclude that
our solution corresponds to a $c=1$ theory.

In order to distinguish between different multi-critical points which are
labeled by $K$, let us calculate the anomalous dimensions of proper
conformal operators. This can be done by calculating their correlation
functions with the puncture operator. For the partial derivative of $W$,
given by Eq.(\ref{4.8}), over $\alpha$, one gets
\be
{\partial \over \partial\alpha }W[\h ] =
 \alpha \log (c-b^2) - c +3 b^2
+ \frac 1N \sum _i  \log{\left(\h_i+2b+\sqrt{\h_i^2+4b\h_i+4c} \,\,\right)}.
\label{5.12}
\ee
It is easy to verify that the r.h.s. is again stationary w.r.t. $b$ and
$c$ due to Eqs.(\ref{5.6}), (\ref{5.7}).

The correlation functions can now be obtained expanding the r.h.s. of
Eq.(\ref{5.12}) in
$\frac{1}{\h_i}$ and representing the result via $t_n$'s.
The exact answer reads
\be
{\partial \over \partial\alpha }W[\h ] =\frac 1N \sum _i  \log{2\h_i}+
 \alpha \log (c-b^2) - \sum_{n=1}^{\infty}\frac 1n t_n
P_{n}\left(-\frac {b}{\sqrt c}\right) \left( 2\sqrt c \right)^{n}.
\label{5.13}
\ee
We have explicitly verified this formula up to $n=4$. The proof can be
done by showing that after differentiation w.r.t.\ $b$ and $c$ and using
the known rules for the derivative of Legendre polynomials:
\be
(1-x^2)P_n^\prime (x) = n[P_{n-1}(x)-xP_n(x)],
\ee
one obtains Eqs.(\ref{5.6}), (\ref{5.7}).

The first term on the r.h.s.\ of Eq.(\ref{5.13}) is due to an extra
factor in Eq.(\ref{2.3}) in front of the integral while the remaining part
looks very similar to the corresponding representation
of the derivative of the free energy of the Kontsevich model w.r.t. the
cosmological constant in genus zero (see, e.g. \cite{Mak91}). Now we have
two potentials $b$ and $c$ so that functions of the ratio
$\frac {b}{\sqrt c}$ appear which are given by Legendre polynomials.

The correlators of proper conformal operators ${\cal O}_n $ with the puncture
operator ${\cal P}= {\cal O}_0$
can now be obtained by differentiating Eq.(\ref{5.13}). One gets
\be
\left\langle {\cal O}_n{\cal P}\right\rangle \equiv
{\partial^2 \over \partial t_n \partial\alpha }W[\h ] =
-\frac 1n
P_{n}\left(-\frac {b}{\sqrt c}\right) \left( 2\sqrt c \right)^{n}.
\label{5.14}
\ee
This formula is an analog of the spherical limit of Gelfand--Dikii
differential polynomials in the case of the Kontsevich model.
For the limiting case (i) above, only correlators with even $n=2m$ are
nonvanishing and one obtains
\be
\left\langle {\cal O}_{2n}{\cal P}\right\rangle =
(-)^{m+1}{(2m-1)!\over (m!)^2}\alpha^m
\ee
in an agreement with an explicit calculation of Ref.\cite{CP91}.

Let us now return to our solution (\ref{4.16}), (\ref{4.17}). One sees
that while the correlator of two puncture operators, given by
Eq.(\ref{4.11}), depends on $\alpha$ logarithmically, the operator
${\cal O}_n$
is scaled as $\langle {\cal O}_n \rangle \sim \alpha^{\frac nK}$.
This dependence of $n$ agrees with the KPZ \cite{KPZ88} spectrum
for $c=1$.

We think that further studies of the genus zero solutions to the model
(\ref{2.1}) as well as their extensions to higher genera deserve future
investigation.
\setcounter{section}{0}
\eop

\appendix{\ \ \ \ Explicit solution at large N}

\setcounter{equation}{0}
Eq.(\ref{4.5}) can be explicitly solved as $N\rightarrow \infty$ while
$\h _i=O(1)$. In this standard planar approximation all terms in the
exponent (\ref{2.5}) (as well as $\log{\Delta[x]}$) are of the same order
$O(N^2)$. The solution (\ref{4.8}) may be obtained along the line
of the method of Br\'ezin and Gross \cite{BG80} introduced
for the unitary matrix model in an external field and applied to the
hermitian
one-matrix model with a cubic potential in \cite{KK89,GN91,MS91}.

At $N=\infty$ one defines the density of eigenvalues of $\h$:
\be
\rho(x)= \frac 1N \sum _{i=1} ^N \delta (x-\h _i),
\label{A.1}
\ee
so that $W[\h ]$ becomes a functional of $\rho$ while
\be
\frac 1N \left.{\partial W[\h ] \over \partial{\h _i}} \right|_{\h _i=x} =
{d\over dx} {\delta W[\rho] \over \delta \rho(x)} \equiv W(x).
\label{A.2}
\ee
Noticing that $dW(x)/dx$ is multiplied by $1/N^2$ --- the factorization
property in the large-N limit --- and can be omitted, Eq.(\ref{4.5}) takes the
form of an integral equation
\be
W^2(x)+\int _a ^d dy\rho(y) {W(y)-W(x)\over y-x} +xW(x)=\alpha\quad
\hbox{for \ }x \in [a,d]
\label{A.3}
\ee
where $[a,d]$ is the support of $\rho$.

It is convenient to make a shift $W(x)\rightarrow W(x)-\frac x2$. Then
Eq.(\ref{A.3}) transforms into
\be
W^2(x)+\int _a ^d dy\rho(y) {W(y)-W(x)\over y-x} =\frac {x^2}{4}+\tilde\alpha
\quad\hbox{for \ }x \in [a,d]
\label{A.4}
\ee
where $\tilde\alpha=\alpha+\frac 12$.

To solve Eq.(\ref{A.4}) one reduces it to a Riemann--Hilbert problem. Let us
define two functions
\be
f(z)=\int _a ^d dx{\rho(x)\over z-x}
\label{A.5}
\ee
and
\be
F(z)=\int _a ^d dx{\rho(x) W(x) \over z-x}
\label{A.6}
\ee
which are analytic with cuts from $a$ to $d$. It follows from
Eqs.(\ref{A.5}),(\ref{A.6}) that
\be
\Im F=W(x)\Im f \qquad\hbox{for \ } x\in [-\infty,+\infty].
\label{A.7}
\ee

The idea is to choose $\Re F$ to satisfy Eq.(\ref{A.4}), i.e. to have
\be
\Re F= W^2(x)+W(x)\Re f-\frac {x^2}{4}-\tilde\alpha
\qquad\hbox{for~}x\in [a,d].
\label{A.8}
\ee
This suggests the following analytic ansatz
\be
F(z)=W^2(z)+W(z)f(z)-z^2/4-\tilde\alpha ,
\label{A.9}
\ee
where $W(z)$ is a real analytic function restricted by
\be
\Im W (\Im W+\Im f) =0\qquad \hbox{for~} x \in [a,d]
\label{A.10}
\ee
and
\be
\Im W (\Re f+2\Re W) =0\qquad \hbox{for~} x \in [-\infty,+\infty]
\label{A.11}
\ee
in order to satisfy Eqs.(\ref{A.8}),(\ref{A.9}). The third restriction on
$W(z)$
is imposed by the asymptotic condition
\be
W(z) \longrightarrow_{z \rightarrow \infty} z/2+O(z^{-1})
\label{A.12}
\ee
which is a consequence of Eq.(\ref{A.4}).

To construct $W(z)$ that satisfies (\ref{A.10})--(\ref{A.12}), one assumes
that
$\Im W\neq 0$ for $x\in[-\beta_{-},-\beta_{+}]$, where $-\beta_{+}<a$
to satisfy Eq.(A.10). Let $-\beta_{-}$ and $-\beta_{+}$ are the roots of
equation $\beta ^2/4+b\beta +c=0$.
Now Eq.(\ref{A.11}) implies
\be
\Im \left( W(x) \qqq x \right) = -{1 \over 2} f(x) \sqrt{-\frac{x^2}{4}-bx-c}
\qquad\hbox{for \ } x \in [-\beta_{-},-\beta_{+}]
\label{A.13}
\ee
since $f(x)$ is real on the real axis outside $[a,d]$. Then
Cauchy's theorem unambiguously determines $W(x) \qqq x$ to be
\bea
W(x) \qqq x= {x^2 \over 4}+\frac 12 bx+\frac c2 -{b^2 \over 2}+
{\tilde\alpha \over 2} - \nonumber \\ - {1 \over 2}\int _a ^d dy{\rho(y)
 \over \qqq x}{\qqq y -\qqq x \over y\,-\,x}
\label{A.14}
\eea
where one has used Eq.(\ref{A.5}) and the additive polynomial has  been
determined to satisfy the asymptotic condition (\ref{A.12}).

To fix $b$ and $c$ we impose the
requirement that $W^2(z)$ does not have poles at $z=-\beta_{\pm}$
which have been
assumed in obtaining Eq.(\ref{A.11}) from Eq.(\ref{A.9}). As follows
from Eq.(\ref{A.14}), the poles are eliminated providing
\bea
b+\frac 14 \int _a ^d dy{\rho(y)  \over \qqq y}=0 \nonumber\\
\frac 14 \int _a ^d dy{y\rho(y)  \over \qqq y}+c-3b^2=\tilde\alpha .
\label{A.15}
\eea
Finally, using Eq.(A.15), $W(x)$ can be represented as
\be
W(x) = \qqq x + {1 \over 2}\int _a ^d dy{\rho(y)
 \over \qqq y}\,{\qqq x -\qqq y \over x\,-\,y},
\label{A.16}
\ee
and the absence of undesirable poles is explicit in this form.

This form of solution is also convenient to show of how the obtained $W(x)$
satisfies Eq.(\ref{A.2}). When $W(x)$ given by (\ref{A.16}) is substituted
into the
l.h.s. of Eq.(\ref{A.2}), the terms which do not contain $\rho$ as well as
those
linear and bilinear in $\rho$ emerge. The key point is to show that the
bilinear in $\rho$ term splits into the product of two functionals linear
in $\rho$. Namely, after symmetrization this term became
\be
- \frac {1}{8} \left\lgroup \int _a ^d dy{\rho(y)  \over \qqq y}
\right\rgroup ^2.
\ee
Other terms can be adjusted to eliminate the $\rho$-dependence. At this
point the restrictions (\ref{A.15}) arose.

The obtained $b$ and $c$ are functionals sophistically depended on
$\rho(x)$. To obtain $W[\rho]$ one should integrate $W(x)$ over $x$ and
$\rho$. The simplest way to do it is to find such expression for $W(x)$ which
is stationary w.r.t.\ the partial differentiation over $b$ and $c$. Then we
can obtain $W[\rho]$ since the total derivative in $\rho$ will coincide with
the partial derivative. It appears that the expression (\ref{A.14}) possesses
the necessary property and after a little algebra we obtain $W[\rho]$:
\bea
W[\rho] = f(\aaa)+\frac 12 \left( \aaa -\frac 12\right)^2\log (c-b^2)
-\frac 52 b^2c -\aaa c +\frac {c^2}{4} +3\aaa b^2 +\frac 94 b^4+\nonumber\\
+ \int _a ^d dx\rho(x) \left[\left(\frac x2 -b\right)\qqq x
+\aaa \log \left(x+2b+\sqrt{x^2+4bx+4c} \right)\right] -\nonumber\\
- {1 \over 4}\iint _a ^d dx\, dy \rho(x) \rho(y)
\log{\left[{xy \over 4} +\frac b2 (x+y) +c+\qqq x \qqq y \,\,\, \right]}
\label{A.17}
\eea
where the integration "constant" $f(\aaa)$ is not essential for applications.
Similarly to the strong-coupling solution of Br\'ezin and Gross \cite{BG80},
this $W[\rho]$ is stationary w.r.t. $b$ and $c$.

Coming back to Eq.(\ref{A.3}), the solution to it differs from (\ref{A.17}) by
an irrelevant to the critical behaviour factor $-\frac 14 \int _a ^d
dx\,x^2\rho (x)$.

\end{document}